\documentclass[10pt, conference]{IEEEtran}

\usepackage{booktabs}
\usepackage{soul}
\usepackage{epsfig}
\usepackage{multirow}

\usepackage{cite}
\usepackage{amsmath,amssymb,amsfonts}
\usepackage{algorithmic}
\usepackage{graphicx}
\usepackage{textcomp}
\usepackage{xcolor}
\def\BibTeX{{\rm B\kern-.05em{\sc i\kern-.025em b}\kern-.08em
    T\kern-.1667em\lower.7ex\hbox{E}\kern-.125emX}}

\usepackage{algorithmic}

\ifCLASSOPTIONcompsoc
 \usepackage[caption=false,font=normalsize,labelfont=sf,textfont=sf]{subfig}
\else
 \usepackage[caption=false,font=footnotesize]{subfig}
\fi

\begin{document}

\title{Provisioning for Solar-Powered Base Stations Driven by Conditional LSTM Networks}

\author{\IEEEauthorblockN{Yawen Guo}
\IEEEauthorblockA{Department of Electrical and\\
Computer Engineering\\
University of California, Santa Cruz \\
Email: yguo127@ucsc.edu}

\and
\IEEEauthorblockN{Sonia Naderi}
\IEEEauthorblockA{Department of Electrical and\\
Computer Engineering\\
University of California, Santa Cruz \\
Email: sonaderi@ucsc.edu}

\and
\IEEEauthorblockN{Colleen Josephson}
\IEEEauthorblockA{Department of Electrical and\\
Computer Engineering\\
University of California, Santa Cruz \\
Email: cjosephson@ucsc.edu}

}

%

\maketitle
\begin{abstract}
Solar-powered base stations are a promising approach to sustainable telecommunications infrastructure. However, the successful deployment of solar-powered base stations requires precise prediction of the energy harvested by photovoltaic (PV) panels vs. anticipated energy expenditure in order to achieve affordable yet reliable deployment and operation. This paper introduces an innovative approach to predict energy harvesting by utilizing a novel conditional Long Short-Term Memory (Cond-LSTM) neural network architecture. Compared with LSTM and Transformer models, the Cond-LSTM model reduced the normalized root mean square error (nRMSE) by 69.6\% and 42.7\%, respectively. We also demonstrate the generalizability of our model across different scenarios. The proposed approach would not only facilitate an accurate cost-optimal PV-battery configuration that meets the outage probability requirements, but also help with site design in regions that lack historical solar energy data.
\end{abstract}
\vspace{-0.5em}

%
\IEEEpeerreviewmaketitle

\section{Introduction}
The energy consumption of information and communication technologies (ICT) has been steadily growing and is about 10\% of worldwide electricity consumption today~\cite{gelenbe_electricity_2023}. The burning of fossil fuels significantly contributes to global climate change, leading to substantial biodiversity loss, widespread pollution, and posing a serious threat to the health and stability of planetary ecosystems. Utilizing renewable energy sources is a critical strategy for reducing carbon emissions and the overall carbon footprint. Fortunately, there is a growing trend towards renewable energy sources. 
Solar-powered base stations significantly  reduce carbon emissions, as well as potential costs savings in the long term by avoiding the need to pay for energy. These ``off-the-grid" base stations also have the added benefit of being resilient against natural disasters. Rather than relying on backup diesel generators, solar-powered base stations present a sustainable alternative for temporary or permanent climate-resilient infrastructure. The challenge lies in designing these stations to ensure their reliable operation. This involves a delicate balance between having sufficient solar panels and batteries for continuous power, and minimizing these components to save costs. 

Accurately predicting energy income vs. energy demand is crucial for designing effective solar-powered base stations. Two important design parameters are the number of photovoltaic (PV) cells, and the battery capacity. These parameters are informed by the desired quality of service (guarantees of throughput, latency, uptime, etc.). In this work, we consider uptime, or outage probability,  as the most fundamental aspect of quality of service. If the batteries fail to provide sufficient energy, the base station experiences an outage. The design goal is to maximize the uptime while minimizing the system cost. Accurate prediction of energy income allows us to minimize the number of purchased PV panels/batteries needed to achieve a desired outage probability. 

Different from the prior studies, this work explores a purely solar-powered macro base station, aligning the power consumption model with typical 5G sites. 
This paper introduces the Cond-LSTM model, designed to achieve more precise predictions, particularly benefiting macro base stations, which consume significantly more energy than previously-studied LTE or 5G micro base stations. 

In summary, our contributions consist of:
\begin{enumerate}
\item Creation of a highly-accurate conditional Long Short-Term Memory (Cond-LSTM) neural network to predict the energy income of solar base station sites, and a corresponding algorithm to provision the minimum solar equipment to achieve desired system performance.
\item Adapting existing power consumption models to be appropriate for macro 5G sites.
\item Evaluation of the Cond-LSTM model versus other approaches (Markov, non-conditional LSTM, Transformer) across various geographical regions, as well as evaluation of how well our model generalizes to new sites.
\end{enumerate}
\section{Background}
Many prior works in this area rely on Markov models~\cite{miozzo_solarstat_2014}. Chamola and Sikdar used a Markov process to model the PV panel harvesting energy and battery level to evaluate the outage probability of a solar powered base station~\cite{chamola_resource_2014, chamola_outage_2015}. Gorla and Chamola~\cite{gorla_battery_2021} utilized a Markov process to estimate the overall battery lifetime for a solar powered cellular base station with a given PV panel wattage. \cite{sharma_review_2020, viswanatha_rao_enhancing_2021, dharmaraja_analytical_2022} also introduced Markov chain as a commonly published process for predicting solar radiation and energy harvesting. Markov models are relatively simple to implement and understand, making them an efficient choice for modeling PV panel harvesting. However, they oversimplify complex weather patterns by dividing them into several discrete categories types, leading to limited accuracy.

LSTM models are an alternative approach that utilizes machine learning for predicting energy harvesting, demonstrating notable accuracy~\cite{hossain_short-term_2020,liu_simplified_2021,lim_solar_2022}. Ku et al. \cite{ku_state_2020} utilized a Convolutional Neural Networks (CNN)-LSTM model to predict energy states in mobile edge computing systems. Han et al. \cite{han_optimal_2022} present a prediction method  based on LSTM for the stand-alone photovoltaic/wind/battery microgrid. Our proposed Conditional LSTM (Cond-LSTM) model bifurcates the network into two components: classification and LSTM. The outcome of the classification is then fed into the LSTM network as a conditional input to enhance its performance. Cond-LSTM models have demonstrated their utility in applications such as melody generation~\cite{yu_conditional_2021} and spoken dialogue systems~\cite{wen_semantically_2015}, but to the best of our knowledge, Cond-LSTMs have yet to be explored for energy harvesting predictions. Our work pioneers the application of Cond-LSTM in telecommunications provisioning by predicting energy harvesting from PV panels. Transformer models, which are based solely on attention mechanisms, without using convolutional or recurrent layers \cite{vaswani2017attention}, have emerged since 2019 and demonstrated powerful capabilities across a wide range of modern deep learning applications. Several papers have published applications of Transformer models and hybrid methods in solar energy forecasting, such as \cite{al2023solar, pospichal2022solar,lopez2022application,munsif2023ct}. In the following sections, we compare the performance of our Cond-LSTM model against both Markov, traditional LSTM and Transformer approaches. 


\begin{figure}[!t]
\centering
\includegraphics[width=0.45\textwidth]{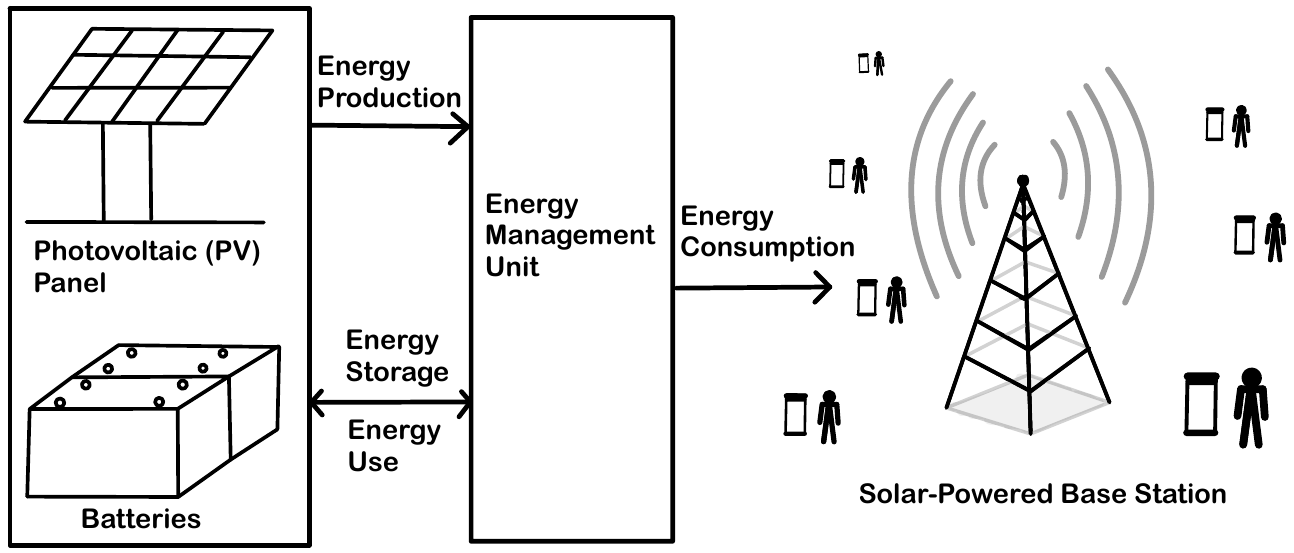}
\vspace{-1.0em}

\caption{Architecture of the Solar-Powered Base Station System.}
\vspace{-1.5em}

\label{Architecture}
\end{figure}

\section{System Model}
Figure \ref{Architecture} presents the high-level architecture of a solar-powered base station. This system harnesses energy from sunlight through PV panels, which, in conjunction with batteries, powers the base station. During sunny conditions, the PV panels generate sufficient energy, a portion of which directly powers the base station, while the excess is stored in batteries. In contrast, during nighttime or cloudy days when the PV panels are unable to generate adequate energy, the energy stored in the batteries is utilized to sustain the operations of the base station. 


\vspace{-0.5em}
\subsection{Harvesting Energy Models}


Traditional LSTM models exhibit satisfactory performance in predicting harvesting energy for micro systems but encounter limitations with macro systems. Transformer models achieve slightly higher accuracy than LSTM\cite{sherozbek2023transformers} but need more calculation resource consumption. Motivated by the challenge of enhancing solar energy prediction accuracy for macro base stations, we introduce the Cond-LSTM model to significantly improve prediction accuracy while having a smaller model size. Specifically, we construct a conditional neural network tailored to solar activity patterns. This is crucial because, for high-capacity PV panels, data during sunlight hours display significant variability compared to other times. Segregating predictions for non-sunlight hours into a separate neural network diminishes data variability and improves data smoothness, thereby facilitating more accurate forecasting. The inputs we are using include time information (Month, Hour), Direct Normal Irradiance (DNI), Diffuse Horizontal Irradiance (DHI), Global Horizontal Irradiance (GHI), Dew Point, Temperature, Pressure, Humidity, Wind Direction, Wind Speed, and Surface Albedo. Differing from a traditional LSTM model, our model instead divides the inputs into two parts. Inputs with selected feature DHI are fed into a self-defined lambda layer, while other features are fed into the LSTM model. DNI, DHI, and GHI are three key features most closely related to solar activity. While DNI is strongly influenced by the angle of sunlight and approaches zero during sunrise and sunset, we consider DHI, which is more associated with diffuse radiation, as the input for the lambda layer. GHI, encompassing both DNI and DHI, could also be used for the lambda layer, but we allocate it to the LSTM layer to aid in prediction. The output of lambda layer will decide if we will accept the outputs from LSTM layer. Table \ref{tab:table_hyperparameter} displays the hyperparameters of LSTM and Cond-LSTM, with the majority kept consistent to ensure a fair comparison, except for a few essential modifications necessitated by the model construction. Both models are single-layer LSTM/Cond-LSTM architectures, each containing one dense layer.

\begin{table}[t]
\caption{Hyperparameters Applied for Cond-LSTM and LSTM .}
\vspace{-1.5em}

\label{tab:table_hyperparameter}
\begin{center}
\resizebox{1.0\columnwidth}{!}{
    \begin{tabular}{c c c}
    \toprule
    Hyperparameter & LSTM & Cond-LSTM  \\ 
    \midrule 
    \midrule 
    \text{Activation} &  Dense & Dense \\
    \text{Input}  & All the Features  &  Feature ``DHI" and other features  \\
    \text{Loss} &  MSE & MSE  \\
    \text{Optimizer} &  Adam & Adam  \\
    \text{Early Stopping} &  Max 200 epochs & Max 200 epochs\\
    \bottomrule
    \end{tabular}
}
\vspace{-2.7em}
\end{center}
\end{table}

We utilize 21 years of statistical weather data provided by the National Renewable Energy Laboratory (NREL)~\cite{noauthor_weather_nodate}, comprising hourly solar irradiance data for specific locations. 
The weather data is then processed through  PySAM of the System Advisor Model (SAM) Software Development Kit ~\cite{noauthor_pysam_nodate} to generate the hourly energy output of a PV panel with a specified rating. This serves as the ground truths for our evaluations. Although we evaluated only four typical locations in different climate zones across the United States, the generalization evaluation in this paper demonstrates that the proposed model, with stable and high accuracy, effectively learns the relationship between weather and solar harvesting. This model can be generalized to other climate zones without retraining and still achieve high prediction accuracy when local weather data is provided.
 
\begin{figure}[!t]
\centering
\includegraphics[width=0.45\textwidth]{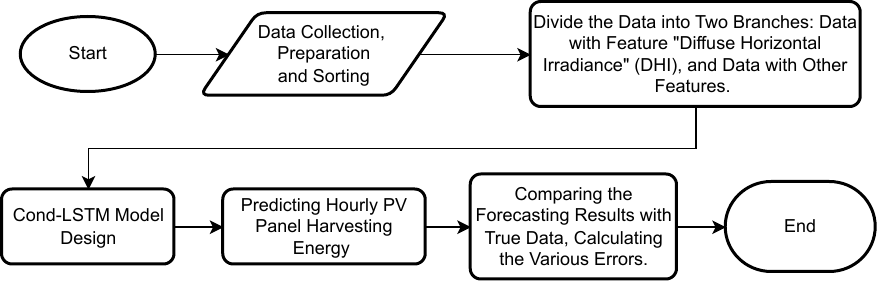}
\vspace{-0.5em}

\caption{Flow Chart of Cond-LSTM Model.}
\label{flowchart}
\vspace{-1.5em}
\end{figure}

In addition to AI/ML-based modeling, 
we also compare our results against a three-state first-order Markov process. The details of this Markov chain can be found in \cite{gorla_battery_2021}.
 In the Markov model, 20 years of data are used to compute the transition matrices for different months, with an additional year of data dedicated to verifying accuracy. For the LSTM, Transformer and Cond-LSTM models, the dataset is segmented into training data (18 years), validation data (2 years), and testing data (1 year). The Cond-LSTM model introduces a conditional neural network architecture designed to process the DHI feature in conjunction with other inputs, effectively distinguishing between non-sunlight hours (with zero harvesting) and sunlight hours. It incorporates LSTM and Dense layers for predictive tasks. Cond-LSTM is applied to the RobustScaler normalization to enhance neural network performance, while we found that LSTMs and Transformers perform better with MinMaxScaler. Furthermore, the training process of all LSTM, Transformer and Cond-LSTM is optimized through the utilization of EarlyStopping, ModelCheckpoint, ReduceLROnPlateau, and a bespoke LearningRateScheduler, all contributing to improved training efficiency and model accuracy. Figure \ref{flowchart} illustrates the process flow for predicting energy harvested by PV panels using the Cond-LSTM model.

\subsection{Traffic Model}

We have grounded our power consumption model in 5G data reported in recent years. Specifically, our traffic model leverages generative normalized traffic data~\cite{hui_large-scale_2023}, and the generative model is itself trained on empirical data. The normalized traffic data has been adjusted to correspond with 5G power consumption as reported by Huawei~\cite{hardesty_5g_2020}. Based on Huawei's data indicating that the Remote Radio Unit (RRU) and Baseband Unit (BBU) requirements for each site require a power supply exceeding 11.5 kilowatts, we assume a peak power consumption of 11.5 kW. 
 The remaining power consumption is calculated using the following formula~\cite{sharma_analysis_2021}:\vspace{-1em}


\vspace{-0em}
\begin{equation*}
P_{bs} = \frac{(N_S \times N_{Tx})(\frac{P_{Tx}}{\eta_{PA}}+\frac{P_{BB}}{\eta_{BB}}+\frac{P_{RF}}{\eta_{RF}})}{(1-(c_4+\epsilon_4n))(1-(c_5+\epsilon_5n))(1-(c_6+\epsilon_6n))}
\end{equation*}

\noindent where the $P_{Tx}$, $P_{BB}$, and $P_{RF}$ are the ideal power consumption for the power amplifier, baseband unit, and transceiver, respectively. They were taken as $P_{Tx} = 500W$, $P_{BB} = 100W$, and $P_{RF} = 100W$. The efficiency $\eta$ of these parameters depends on the number of connections $n$ (load), as follows: $\eta_{PA} = c_1-\epsilon_1n$, $\eta_{BB} = c_2-\epsilon_2n$ and $\eta_{RF} = c_3-\epsilon_3n$ respectively, where $\epsilon_i$ is a coefficient for the efficiency varies with load $n$, and $c_i$  represents the part of the efficiency that is not load-dependent.  $\sigma_{R} = c_4+\epsilon_4n$, $\sigma_{DC} = c_5+\epsilon_5n$ and $\sigma_{COOL} = c_6+\epsilon_6n$ are the loss factors due to the rectification, regulation, and cooling respectively.  All the efficiency and loss factors above are between 0 and 1. The parameter $\epsilon_i$ is contingent upon the hardware design and specifications of the device and is considered negligible unless the load approaches the magnitude of $10^5–10^6$. Peak load is around 14000, so variation of efficiencies with the load is very weak, with all the $\epsilon_i$ were taken as $10^{-7}$. The efficiencies were taken as $c_1 = 0.75$, $c_2 = c_3 = 0.8$, $c_4 = 0.65$, $c_5 = 0.15$, which we take the same parameters as the article \cite{sharma_analysis_2021}. $c_6$ is set to be 0.47 according to \cite{zhou_energy-saving_2013}. Each base station typically has 3 hexagonal sub-sectors (thus $N_S = 3$) and we assume every antenna is deployed for 10,000 devices as in \cite{sharma_analysis_2021}. So the number of antennas for each sub-sector is $N_{Tx} = \frac{n}{10^4}$.
The auxiliary power is neglected because of its independence of the load.


\subsection{Sizing the PV Panels and Battery}

\emph{PV Panel:} We select a configuration of 72 cells\footnote{For both PV panels and batteries, the prices are drawn from the options available in an online retailer as of Mar 19th, 2024. For cost comparisons, we use Mission Solar 430W monocrystalline MIN-MSE430SX9Z (\$244.22) and the Generac PWRcell 3.0kW Lithium-Ion Battery Module G0080040 (\$2093.37).}.

\emph{Battery Bank:} Lithium-ion batteries, due to reliability, modularity, and durability.

\emph{Optimal Sizing:} To determine the optimal number of PV panels and batteries  while considering the tolerable outage probability and the need to minimize costs, we assess the residual energy from the previous hour in the batteries, $E_{Battery}(i-1)$, plus the energy harvested in the current hour, $E_{harvest}(i)$. We then compare this sum to the current hour's energy consumption, $E_{consume}(i)$, where $i$ denotes the hour. We assume that all Lithium-Ion batteries are fully charged prior to the base station's operation, with each battery's capacity, $C_B$, being 3.0 kWh. To prevent battery damage due to excessive discharge, a lower limit is established at 20\% of a battery's capacity. Consequently, the base station operates normally if $m \cdot E_{Battery}(i-1)+n \cdot E_{harvest}(i) - E_{consume}(i) >~0.2 \cdot m \cdot C_B$, where the $m$ is the number of batteries and $n$ is the number of PV panel modules. Conversely, an outage occurs if $m \cdot E_{Battery}(i-1) + n \cdot E_{harvest}(i) - E_{consume}(i) < 0.2 \cdot m \cdot C_B$. In the sizing model, it is assumed that a maximum of one hour of outage is permissible over a span of approximately a month ($len\_data = 671\text{ }hours$), which corresponds to 99.9\% uptime. In this analysis, installation and maintenance expenses are not considered. The constrained optimization problem is  described below:

\noindent\begin{IEEEeqnarray*}{rCl}
    \text{Minimize } \texttt{{\small cost}} & = & n \cdot \texttt{{\small PV\_cost}} + m \cdot \texttt{{\small Battery\_cost}},\\
    \text{subject to }
    \vspace{-1em}
     E_{\text{Battery}}(0) & = & m \cdot C_B. \nonumber\\
\end{IEEEeqnarray*}



\text{Available Energy:}
\vspace{-0.5em}
\begin{IEEEeqnarray*}{rCl}
\hspace{-1em}    E_{\text{avail}}(i) & = & E_{\text{Battery\_trim}}(i-1) + n \cdot E_{\text{harvest}}(i) - E_{\text{consume}}(i)
\end{IEEEeqnarray*}

\text{Outage Constraints:}
\vspace{-0.5em}
\begin{IEEEeqnarray*}{rCl}
 E_{\text{avail}}(i) & \geq & 0.2 \cdot m \cdot C_B  + \varepsilon - M \cdot \text{outage\_indice}(i)\\
    E_{\text{avail}}(i) & \leq & 0.2 \cdot m \cdot C_B \nonumber + \varepsilon + M \cdot (1 - \text{outage\_indice}(i))
\end{IEEEeqnarray*}

\text{Battery Energy Update:}
\vspace{-0.5em}
\begin{IEEEeqnarray*}{rCl}
    E_{\text{Battery}}(i) & = & E_{\text{avail}}(i) + E_{\text{consume}}(i) \cdot \text{outage\_indice}(i)
\end{IEEEeqnarray*}

\text{Trimmed Battery Constraints:}
\vspace{-0.5em}
\begin{IEEEeqnarray*}{rCl}
  E_{\text{Battery\_trim}}(i) & \leq & 
    \min\left( E_{\text{Battery}}(i), m \cdot C_B \right)
\end{IEEEeqnarray*}

\text{Battery Capacity Constraints:}
\vspace{-0.5em}
\begin{IEEEeqnarray*}{rCl}
    E_{\text{Battery}}(i) & \leq & m \cdot C_B \nonumber \\
    E_{\text{Battery}}(i) & \geq & 0
\end{IEEEeqnarray*}
\vspace{-2.7em}

\begin{IEEEeqnarray*}{rCl}
\text{Total Outage Constraint:}
    \sum_{i=1}^{\text{len\_data}} \text{outage\_indice}(i) & \leq & 0
\end{IEEEeqnarray*}

A Mixed-Integer Linear Programming (MILP) approach with big-M method is used to solve this problem, where M is a sufficiently large constant to force a constraint to be either active or inactive. Taking the Iowa (IA) region as an example, for 99.9\% uptime, the harvesting data are calculated via Cond-LSTM. Then the data are used in the MILP to solve the optimal solutions for the number of solar panels and storage batteries 
(which are 47 and 22 respectively for the Iowa (IA) in our simulation). 
As the day progresses, the battery bank becomes fully charged to a 66 kWh capacity, which corresponds to 52.8 kWh of usable capacity due to the system policy of never draining the batteries below 20\%. 



\section{Evaluations}
\vspace{-0.25em}

\subsection{Evaluation: Energy Harvesting Forecasting Methods}
\vspace{-0.25em}
In this section, we evaluate the energy harvesting predictions made by a Cond-LSTM, in comparison LSTM , Transformer and Markov models. Table~\ref{table_different_methods} assesses the models' accuracy for a 48.94 kW panel through RMSE for hourly and daily predictions, as well as Mean Absolute Error (MAE), Mean Error (ME), and Mean Percentage Error (MPE) for daily energy harvesting predictions. The results quantify the error metrics in watts, where `H' denotes hourly predictions and `D' indicates daily estimates. An arrow pointing downward signifies that lower values reflect superior predictive performance. The comparative analysis reveals that the Cond-LSTM model significantly outperforms the other approaches.  
We used the same dataset for all models evaluated in this paper to ensure fairness. Variations in PV panel sizes and data lengths can significantly impact most error metrics. Normalized RMSE\footnote{nRMSE is defined as $nRMSE = RMSE/(A_{max} - A_{min})$, where A represents actual data} (nRMSE) is provided for comparing with results from other papers, as this metric is not significantly affected by PV panel size or data length. Prior works found that LSTMs achieve 3.6\% nRMSE with 600 epochs\cite{obiora_forecasting_2020}. 
In comparison, our Cond-LSTM model achieves 0.78\% nRMSE with a maximum of 500 epochs.



\begin{table}[!t]
\caption{Cond-LSTM vs. traditional LSTM, Transformer and Markov for a 48.94kW rating PV Panel.}
\label{table_different_methods}
\centering
\resizebox{1.0\columnwidth}{!}{
    \begin{tabular}{|c|c|c|c|c|c|c|}
    \hline
    Methods     & \begin{tabular}[c]{@{}c@{}}nRMSE (H)$\downarrow$\\ (\%)\end{tabular}  &
    \begin{tabular}[c]{@{}c@{}}RMSE (H)$\downarrow$\\ (W)\end{tabular} & \begin{tabular}[c]{@{}c@{}}RMSE (D)$\downarrow$ \\ (W)\end{tabular}  & \begin{tabular}[c]{@{}c@{}}MAE (D)$\downarrow$ \\ (W)\end{tabular}  & \begin{tabular}[c]{@{}c@{}}ME (D)$\downarrow$  \\ (W)\end{tabular}   & \begin{tabular}[c]{@{}c@{}}MPE (D)$\downarrow$  \\ (\%)\end{tabular}  \\ \hline
    Markov   &21.238   & 8588.413 & 120815.054 & 94364.946 & -10471.987 & 52.811  \\ \hline
    LSTM   &2.377     & 1043.646 & 9308.663   & 6990.764  & 2763.534   & 1.932  \\ \hline
    Transformer & 1.480 & 648.136  & 12365.747   & 11448.602  & -11087.669   & -8.342 \\ \hline
    Cond-LSTM & \textbf{0.776} & \textbf{343.856}  & \textbf{2795.262}   & \textbf{2231.747}  & \textbf{-108.615}   & \textbf{-0.410} \\ \hline
    \end{tabular}
}
\vspace{-1.5em}

\end{table}

\begin{figure}[!t] 
\centering
\includegraphics[width=0.45\textwidth]{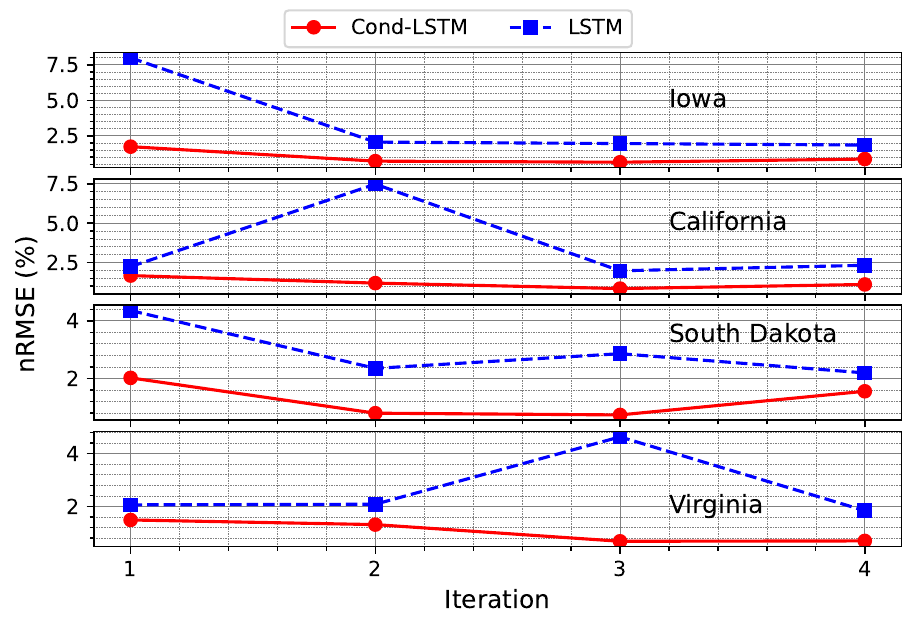}
\label{fig:4fold_nRMSE}
\vspace{1pt}
\vspace{-1em}

\caption{4-fold Cross-Validation Results}
\label{4fold}
\vspace{-1.5em}

\end{figure}

We also evaluated the LSTM and Cond-LSTM models using a time-series split to separate training and testing data across several partitions, allowing us to validate the model predictability across different scenarios. Data were partitioned four ways, with the training and test data distributions set as 4 years/4 years, 8 years/4 years, 12 years/4 years, and 16 years/4 years. 
We applied the time-series split separately for data from four regions, assessing both RMSE and nRMSE for each scenario. 
Fig. \ref{4fold} depicts the time-series split, cross-validation across different regions shows that Cond-LSTM outperforms the traditional LSTM across all four regions.

\vspace{-0.5em}
\subsection{Evaluation: Applicability to Different Regions}
\begin{figure}[!t]
\centering
\includegraphics[width=0.37\textwidth]{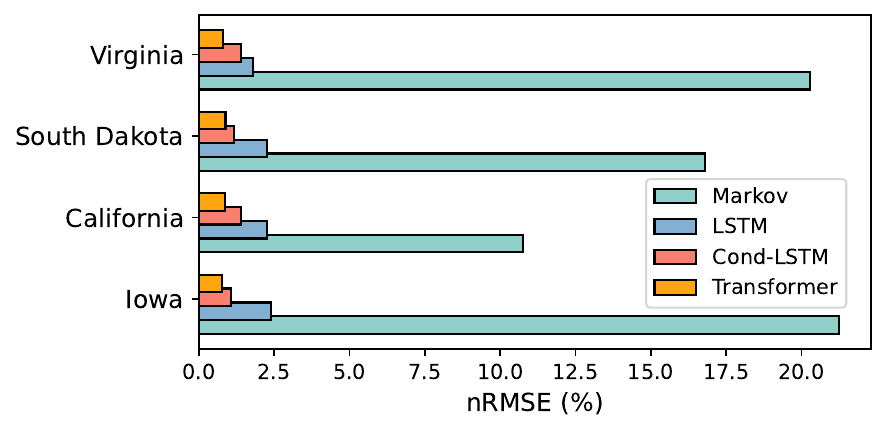}
\vspace{-1em}
\caption {nRMSE of Markov, LSTM, Transformer and Cond-LSTM }
\label{Cond_LSTM_locations2_nRMSE}
\vspace{-1.5em}
\end{figure}

\vspace{-0.25em}
To assess how Cond-LSTM performance changes depending on local weather patterns, we selected several areas characterized by their distinct climatic features and trained a model for each location. We chose Rapid City in South Dakota (SD) (renowned for its unpredictable weather patterns~\cite{silver_which_2014}) to test the model's performance in a variable climate environment. 
Another two regions selected were: California (CA), representing the western United States, and Virginia (VA), reflecting the eastern United States. Similarly, for each location, 20 years of historical data are used for model development and 1 year of data for evaluation. Results show that the RMSE of the Cond-LSTM prediction for different regions still remain low even for a large size PV panel ($48.94 kW$ DC rating). The RMSE of South Dakota was the highest, at $405.973 W$. The RMSEs of all other locations are actually better ($4.39\%-18.45\%$ improvement). The results shows that for all those regions, Cond-LSTM have a stabler and much more accurate prediction than Markov, LSTM and Transformer method.  Figure \ref{Cond_LSTM_locations2_nRMSE} presents the nRMSE (normalized by range) for the Markov, LSTM, Transformer and Cond-LSTM models across four distinct locations. LSTM, Transformer and Cond-LSTM models exhibit stability across varying weather conditions, in contrast to the Markov model, which exhibits instability across different datasets. Notably, the Cond-LSTM achieves the highest prediction accuracy, with an $nRMSE = 0.776\%, 0.861\%, 0.882\%, 0.802\% $ for Iowa, California, South Dakota and Virginia respectively. In addition, we investigate the model sizes of the three machine learning models, as shown in Table \ref{Model_size}. Although the Transformer model achieves comparable accuracy, it has a significantly larger model size in terms of parameters, overall model size, and floating point operations per second (FLOPs). The proposed Cond-LSTM outperforms in both accuracy and resource efficiency.
\vspace{0.01em}


\vspace{-1.5em}

\begin{table}[!t]
\caption{Resource Consumption of Different Models}
\vspace{-0.5em}
\label{Model_size}
\centering

\resizebox{0.7\columnwidth}{!}{
    \begin{tabular}{|c|c|c|c|}
    
    \hline
    \begin{tabular}[c]{@{}c@{}}Methods\end{tabular} & \begin{tabular}[c]{@{}c@{}}Parameter size$\downarrow$\end{tabular} & \begin{tabular}[c]{@{}c@{}}Model size$\downarrow$\end{tabular} & \begin{tabular}[c]{@{}c@{}}FLOPs $\downarrow$\end{tabular} \\ \hline
    LSTM& 5793  & 22.63 KB & 135936  \\ \hline
    Transformer & 118081 & 461.25 KB & 4889088 \\ \hline
    Cond-LSTM & \textbf{5665} & \textbf{22.13 KB} & \textbf{398592}  \\ \hline
    \end{tabular}
}
\vspace{-2.2em}
\end{table}
\vspace{0.0em}
\subsection{Evaluation: Generalizability}
\vspace{-0.7em}
In this section, we evaluate how well a model trained for one location can perform at a new location without re-training. We trained a general model by using 2000 - 2019 data from the Iowa region, and then used the pre-trained model to predict the 2020 year harvesting data for locations in California, South Dakota and Virginia. 
The ability to use a pre-trained model helps avoid the financial and environmental costs of re-training a model for every new location.

The nRMSE (normalized by range) of the general model's prediction for those regions are $2.624\%$, $2.010\%$, $1.162\%$, respectively. Compared to the Cond-LSTM models trained specifically for a region, a non-conditional LSTM trained on local data has a $0.999\% - 1.406\%$ nRMSE difference, while Markov based on the same data has an $9.895\% - 19.466\%$ nRMSE difference. The general Cond-LSTM model, however, only shows $0.360\% - 1.763\% $ nRMSE difference, which is similar to the accuracy of LSTMs specifically trained on the local data. These results extend beyond the specific case of a model trained on Iowa being evaluated in other locations. If the model is trained solely on data from CA, SD or VA, and then evaluated in new locations, the results are similar: the `general' model only has an 2.9\% increase in nRMSE, on average. The results shows that by using a Cond-LSTM, we can use a model trained in one place to predict the harvesting energy of other locations by feeding the weather data to the model with a high accuracy. 

Table~\ref{onetomany_ThreeLocations} presents the PV/battery numbers and associated costs obtained using a model that makes predictions in new locations. Additionally, we compare the performance of Cond-LSTM, Transformer (generalized), and LSTM (generalized) models, demonstrating that the Cond-LSTM predictions closely match the ground truth. This indicates that a model trained with data from one location can still accurately predict the size of a solar-powered system in another. Furthermore, when provisioning for new locations—particularly those with insufficient historical
weather data for effective training or no solar harvesting
data—short-term weather data can be used effectively by leveraging a well-trained model from regions with abundant historical weather and harvesting data.

\vspace{-0.5em}
\subsection{Evaluation: Sizing the Solar Energy System and Cost}
The results in Table \ref{table_sizing} indicate that the Markov models exhibit significant errors in estimating the size of solar-powered systems. Although the LSTM model and Transformer yields good results, the Cond-LSTM model outperforms both. However, this marginal discrepancy between the LSTM, Transformer and Cond-LSTM models can be attributed to the low DC rating (0.43 kW) used for a single PV panel module. The benefits of our algorithm become more pronounced when using larger PV panels or custom-designed integrated panels. 
As the size of each panel grows, over or under provisioning by a single panel can cause the error to amplify in size due to the increased power per PV panel. This will lead to considerable increases in RMSE for the non Cond-LSTM approaches, as demonstrated in Table \ref{table_different_methods}.

\begin{table}[!t]
\caption{Optimal solar sizing/cost per different prediction models}
\vspace{-0.5em}
\label{table_sizing}
\centering
\resizebox{0.9\columnwidth}{!}{
    \begin{tabular}{|c|c|c|c|}
    \hline
    \begin{tabular}[c]{@{}c@{}}Harvesting\\ Data\end{tabular} & \begin{tabular}[c]{@{}c@{}}Number of \\ PV Panel Modules\end{tabular} & \begin{tabular}[c]{@{}c@{}}Number of \\ Battery Modules\end{tabular} & \begin{tabular}[c]{@{}c@{}}Cost Difference $\downarrow$\end{tabular} \\ \hline
    Ground Truth & 48 & 22 & 0 (\$57776.70)  \\ \hline
    Markov & 69 & 19 & 1.993\%  \\ \hline
    LSTM & 50 & 22 & 0.845\% \\ \hline
    Transformer & 49 & 23 & 4.046\% \\ \hline
    Cond-LSTM & 47 & 22 & \textbf{0.423\%}  \\ \hline
    \end{tabular}
}
\vspace{-2em}
\end{table}

\vspace{-1.5em}
\begin{table}[!t]
\caption{Optimal sizing and cost of a general model applied to South Dakota, California and Virginia regions}
\vspace{-0.5em}
\label{onetomany_ThreeLocations}
\centering
\resizebox{0.9\columnwidth}{!}{
    \begin{tabular}{|c|ccc|}
    \hline
    \begin{tabular}[c]{@{}c@{}}Harvesting\\ Data\end{tabular} & \multicolumn{1}{c|}{\begin{tabular}[c]{@{}c@{}}Number of\\ PV Panel Modules\end{tabular}} & \multicolumn{1}{c|}{\begin{tabular}[c]{@{}c@{}}Number of\\ Battery Modules\end{tabular}} & \begin{tabular}[c]{@{}c@{}}Cost Difference $\downarrow$\end{tabular} \\ \hline
    Location & \multicolumn{3}{c|}{South Dakota} \\ \hline
    Ground Truth & \multicolumn{1}{c|}{57} & \multicolumn{1}{c|}{21} & 0 (\$57881.31)  \\ \hline
    LSTM & \multicolumn{1}{c|}{58} & \multicolumn{1}{c|}{23} & 7.655\% \\ \hline
    Transformer & \multicolumn{1}{c|}{58} & \multicolumn{1}{c|}{22} & 4.039\% \\ \hline
    Cond-LSTM & \multicolumn{1}{c|}{57}  & \multicolumn{1}{c|}{21} & \textbf{0\%}   \\ \hline
    Location & \multicolumn{3}{c|}{California} \\ \hline
    Ground Truth & \multicolumn{1}{c|}{50} & \multicolumn{1}{c|}{22} & 0 (\$58265.14) \\ \hline
    LSTM & \multicolumn{1}{c|}{51} & \multicolumn{1}{c|}{21} & 3.174\% \\ \hline
    Transformer & \multicolumn{1}{c|}{55} & \multicolumn{1}{c|}{22} & 2.096\% \\ \hline
    Cond-LSTM & \multicolumn{1}{c|}{51} & \multicolumn{1}{c|}{22} & \textbf{0.419\%} \\ \hline
    Location & \multicolumn{3}{c|}{Virginia} \\ \hline
    Ground Truth & \multicolumn{1}{c|}{86} & \multicolumn{1}{c|}{35} & 0 (\$94270.87) \\ \hline
    LSTM & \multicolumn{1}{c|}{86} & \multicolumn{1}{c|}{32} & 6.662\% 
    \\ \hline
    Transformer & \multicolumn{1}{c|}{87} & \multicolumn{1}{c|}{36} & 2.480\% \\ \hline
    Cond-LSTM & \multicolumn{1}{c|}{86} & \multicolumn{1}{c|}{35} & \textbf{0\%} \\ \hline
    \end{tabular}
}
\vspace{-2em}
\end{table}
\vspace{1em}
\section{Conclusion and Future Work}
\vspace{-0.5em}
This paper introduces an accurate forecasting approach Cond-LSTM that designed for optimizing the provisioning of solar-powered macro cellular base stations. Firstly, the performance of the Cond-LSTM model is evaluated against three established forecasting methods: the Markov model, LSTM, and Transformer. Using a comprehensive set of error evaluation metrics, the results demonstrate Cond-LSTM's superior accuracy and smaller model size. Secondly, the versatility of our model across various geographic regions is assessed, demonstrating its broad applicability. Then, incorporating scenarios that closely mirror the energy consumption patterns of macro 5G base stations and a given tolerable power outage rate, we simulated the number of PV panel modules and battery modules required for a solar-powered 5G base station. The results showed that our model is highly similar with the results calculated by true data, enabling precise determination of the required quantities of PV panels and storage batteries. Finally, we show that our model, once trained on data from a single region, can effectively forecast solar-powered system outputs and inform sizing decisions in various other regions. This capability is particularly beneficial for establishing solar base stations in locales without extensive historical data on solar energy capture. In practical applications, the model can be trained using historical solar harvesting data from one area and then applied to forecast solar energy production in regions with only short-term weather data, thereby facilitating the establishment of solar-powered base stations in such areas. 

In the future, the precision of this model will also facilitate the power allocation process, as accurate solar energy forecasting plays a crucial role in the operation of solar-powered base stations. We plan to integrate reinforcement learning in the model for more complex scenarios, such as addressing the impact of shadows in solar energy harvesting, as well as optimizing energy management and resource allocation. Our proposed model, with its smaller size, is more sustainable for applications in smart energy management and resource allocation. We will also consider optimizing the model to further reduce computational resource consumption.

\vspace{-0.5em}

\bibliographystyle{IEEEtran}
\bibliography{BSCondLSTM}

\begin{thebibliography}{10}
\providecommand{\url}[1]{#1}
\csname url@samestyle\endcsname
\providecommand{\newblock}{\relax}
\providecommand{\bibinfo}[2]{#2}
\providecommand{\BIBentrySTDinterwordspacing}{\spaceskip=0pt\relax}
\providecommand{\BIBentryALTinterwordstretchfactor}{4}
\providecommand{\BIBentryALTinterwordspacing}{\spaceskip=\fontdimen2\font plus
\BIBentryALTinterwordstretchfactor\fontdimen3\font minus \fontdimen4\font\relax}
\providecommand{\BIBforeignlanguage}[2]{{%
\expandafter\ifx\csname l@#1\endcsname\relax
\typeout{** WARNING: IEEEtran.bst: No hyphenation pattern has been}%
\typeout{** loaded for the language `#1'. Using the pattern for}%
\typeout{** the default language instead.}%
\else
\language=\csname l@#1\endcsname
\fi
#2}}
\providecommand{\BIBdecl}{\relax}
\BIBdecl

\bibitem{gelenbe_electricity_2023}
\BIBentryALTinterwordspacing
E.~Gelenbe, ``\BIBforeignlanguage{en}{Electricity {Consumption} by {ICT}: {Facts}, trends, and measurements},'' \emph{\BIBforeignlanguage{en}{Ubiquity}}, vol. 2023, no. August, pp. 1--15, Aug. 2023. [Online]. Available: \url{https://dl.acm.org/doi/10.1145/3613207}
\BIBentrySTDinterwordspacing

\bibitem{miozzo_solarstat_2014}
\BIBentryALTinterwordspacing
M.~Miozzo, D.~Zordan, P.~Dini, and M.~Rossi, ``{SolarStat}: {Modeling} photovoltaic sources through stochastic {Markov} processes,'' in \emph{2014 {IEEE} {International} {Energy} {Conference} ({ENERGYCON})}, May 2014, pp. 688--695. [Online]. Available: \url{https://ieeexplore.ieee.org/abstract/document/6850501}
\BIBentrySTDinterwordspacing

\bibitem{chamola_resource_2014}
\BIBentryALTinterwordspacing
V.~Chamola and B.~Sikdar, ``Resource provisioning and dimensioning for solar powered cellular base stations,'' in \emph{2014 {IEEE} {Global} {Communications} {Conference}}, Dec. 2014, pp. 2498--2503, iSSN: 1930-529X. [Online]. Available: \url{https://ieeexplore.ieee.org/abstract/document/7037183}
\BIBentrySTDinterwordspacing

\bibitem{chamola_outage_2015}
------, ``Outage estimation for solar powered cellular base stations,'' in \emph{2015 IEEE ICC}, 2015, pp. 172--177.

\bibitem{gorla_battery_2021}
P.~Gorla and V.~Chamola, ``Battery lifetime estimation for energy efficient telecommunication networks in smart cities,'' \emph{Sustainable Energy Technologies and Assessments}, vol.~46, p. 101205, Aug. 2021.

\bibitem{sharma_review_2020}
\BIBentryALTinterwordspacing
A.~Sharma and A.~Kakkar, ``\BIBforeignlanguage{en}{A review on solar forecasting and power management approaches for energy-harvesting wireless sensor networks},'' \emph{\BIBforeignlanguage{en}{International Journal of Communication Systems}}, vol.~33, no.~8, p. e4366, 2020, \_eprint: https://onlinelibrary.wiley.com/doi/pdf/10.1002/dac.4366. [Online]. Available: \url{https://onlinelibrary.wiley.com/doi/abs/10.1002/dac.4366}
\BIBentrySTDinterwordspacing

\bibitem{viswanatha_rao_enhancing_2021}
S.~Viswanatha~Rao, S.~S. Pillai, and G.~Shiny, ``\BIBforeignlanguage{en}{Enhancing the {Performance} of an {Energy} {Harvesting} {Wireless} {Sensor} {Node} {Using} {Markov} {Decision} {Process}},'' in \emph{\BIBforeignlanguage{en}{Advances in {Electrical} and {Computer} {Technologies}}}, T.~Sengodan, M.~Murugappan, and S.~Misra, Eds.\hskip 1em plus 0.5em minus 0.4em\relax Singapore: Springer Nature, 2021, pp. 581--593.

\bibitem{dharmaraja_analytical_2022}
\BIBentryALTinterwordspacing
S.~Dharmaraja, A.~Aggarwal, and K.~Naresa, ``Analytical {Modelling} and {Simulation} {Of} {DRX} {Mechanism} for {Energy} {Harvesting},'' in \emph{2022 {Annual} {Reliability} and {Maintainability} {Symposium} ({RAMS})}, Jan. 2022, pp. 1--5, iSSN: 2577-0993. [Online]. Available: \url{https://ieeexplore.ieee.org/document/9893979}
\BIBentrySTDinterwordspacing

\bibitem{hossain_short-term_2020}
\BIBentryALTinterwordspacing
M.~S. Hossain and H.~Mahmood, ``Short-{Term} {Photovoltaic} {Power} {Forecasting} {Using} an {LSTM} {Neural} {Network} and {Synthetic} {Weather} {Forecast},'' \emph{IEEE Access}, vol.~8, pp. 172\,524--172\,533, 2020, conference Name: IEEE Access. [Online]. Available: \url{https://ieeexplore.ieee.org/abstract/document/9200614}
\BIBentrySTDinterwordspacing

\bibitem{liu_simplified_2021}
\BIBentryALTinterwordspacing
C.-H. Liu, J.-C. Gu, and M.-T. Yang, ``A {Simplified} {LSTM} {Neural} {Networks} for {One} {Day}-{Ahead} {Solar} {Power} {Forecasting},'' \emph{IEEE Access}, vol.~9, pp. 17\,174--17\,195, 2021, conference Name: IEEE Access. [Online]. Available: \url{https://ieeexplore.ieee.org/abstract/document/9333638}
\BIBentrySTDinterwordspacing

\bibitem{lim_solar_2022}
\BIBentryALTinterwordspacing
S.-C. Lim, J.-H. Huh, S.-H. Hong, C.-Y. Park, and J.-C. Kim, ``\BIBforeignlanguage{en}{Solar {Power} {Forecasting} {Using} {CNN}-{LSTM} {Hybrid} {Model}},'' \emph{\BIBforeignlanguage{en}{Energies}}, vol.~15, no.~21, p. 8233, Jan. 2022, number: 21 Publisher: Multidisciplinary Digital Publishing Institute. [Online]. Available: \url{https://www.mdpi.com/1996-1073/15/21/8233}
\BIBentrySTDinterwordspacing

\bibitem{ku_state_2020}
\BIBentryALTinterwordspacing
Y.-J. Ku, S.~Sapra, S.~Baidya, and S.~Dey, ``State of {Energy} {Prediction} in {Renewable} {Energy}-driven {Mobile} {Edge} {Computing} using {CNN}-{LSTM} {Networks},'' in \emph{2020 {IEEE} {Green} {Energy} and {Smart} {Systems} {Conference} ({IGESSC})}, Nov. 2020, pp. 1--7, iSSN: 2640-0138. [Online]. Available: \url{https://ieeexplore.ieee.org/abstract/document/9285102}
\BIBentrySTDinterwordspacing

\bibitem{han_optimal_2022}
\BIBentryALTinterwordspacing
H.~Han, H.~Liu, X.~Zuo, G.~Shi, Y.~Sun, Z.~Liu, and M.~Su, ``Optimal {Sizing} {Considering} {Power} {Uncertainty} and {Power} {Supply} {Reliability} {Based} on {LSTM} and {MOPSO} for {SWPBMs},'' \emph{IEEE Systems Journal}, vol.~16, no.~3, pp. 4013--4023, Sep. 2022, conference Name: IEEE Systems Journal. [Online]. Available: \url{https://ieeexplore.ieee.org/abstract/document/9681616}
\BIBentrySTDinterwordspacing

\bibitem{yu_conditional_2021}
\BIBentryALTinterwordspacing
Y.~Yu, A.~Srivastava, and S.~Canales, ``Conditional {LSTM}-{GAN} for {Melody} {Generation} from {Lyrics},'' \emph{ACM Transactions on Multimedia Computing, Communications, and Applications}, vol.~17, no.~1, pp. 35:1--35:20, Apr. 2021. [Online]. Available: \url{https://dl.acm.org/doi/10.1145/3424116}
\BIBentrySTDinterwordspacing

\bibitem{wen_semantically_2015}
\BIBentryALTinterwordspacing
T.-H. Wen, M.~Gasic, N.~Mrksic, P.-H. Su, D.~Vandyke, and S.~Young, ``Semantically {Conditioned} {LSTM}-based {Natural} {Language} {Generation} for {Spoken} {Dialogue} {Systems},'' Aug. 2015, arXiv:1508.01745 [cs]. [Online]. Available: \url{http://arxiv.org/abs/1508.01745}
\BIBentrySTDinterwordspacing

\bibitem{vaswani2017attention}
A.~Vaswani, ``Attention is all you need,'' \emph{Advances in Neural Information Processing Systems}, 2017.

\bibitem{al2023solar}
E.~M. Al-Ali, Y.~Hajji, Y.~Said, M.~Hleili, A.~M. Alanzi, A.~H. Laatar, and M.~Atri, ``Solar energy production forecasting based on a hybrid cnn-lstm-transformer model,'' \emph{Mathematics}, vol.~11, no.~3, p. 676, 2023.

\bibitem{pospichal2022solar}
J.~Posp{\'\i}chal, M.~Kubov{\v{c}}{\'\i}k, and I.~Dirgov{\'a}~Lupt{\'a}kov{\'a}, ``Solar irradiance forecasting with transformer model,'' \emph{Applied Sciences}, vol.~12, no.~17, p. 8852, 2022.

\bibitem{lopez2022application}
M.~L{\'o}pez~Santos, X.~Garc{\'\i}a-Santiago, F.~Echevarr{\'\i}a~Camarero, G.~Bl{\'a}zquez~Gil, and P.~Carrasco~Ortega, ``Application of temporal fusion transformer for day-ahead pv power forecasting,'' \emph{Energies}, vol.~15, no.~14, p. 5232, 2022.

\bibitem{munsif2023ct}
M.~Munsif, M.~Ullah, U.~Fath, S.~U. Khan, N.~Khan, and S.~W. Baik, ``Ct-net: A novel convolutional transformer-based network for short-term solar energy forecasting using climatic information.'' \emph{Computer Systems Science \& Engineering}, vol.~47, no.~2, 2023.

\bibitem{sherozbek2023transformers}
J.~Sherozbek, J.~Park, M.~S. Akhtar, and O.-B. Yang, ``Transformers-based encoder model for forecasting hourly power output of transparent photovoltaic module systems,'' \emph{Energies}, vol.~16, no.~3, p. 1353, 2023.

\bibitem{noauthor_weather_nodate}
\BIBentryALTinterwordspacing
``Weather {Data} - {System} {Advisor} {Model} - {SAM}.'' [Online]. Available: \url{https://sam.nrel.gov/weather-data.html}
\BIBentrySTDinterwordspacing

\bibitem{noauthor_pysam_nodate}
\BIBentryALTinterwordspacing
``{PySAM} — {NREL}-{PySAM} 5.1.0 documentation.'' [Online]. Available: \url{https://nrel-pysam.readthedocs.io/en/main/}
\BIBentrySTDinterwordspacing

\bibitem{hui_large-scale_2023}
\BIBentryALTinterwordspacing
S.~Hui, H.~Wang, T.~Li, X.~Yang, X.~Wang, J.~Feng, L.~Zhu, C.~Deng, P.~Hui, D.~Jin, and Y.~Li, ``\BIBforeignlanguage{en}{Large-scale {Urban} {Cellular} {Traffic} {Generation} via {Knowledge}-{Enhanced} {GANs} with {Multi}-{Periodic} {Patterns}},'' in \emph{\BIBforeignlanguage{en}{Proceedings of the 29th {ACM} {SIGKDD} {Conference} on {Knowledge} {Discovery} and {Data} {Mining}}}.\hskip 1em plus 0.5em minus 0.4em\relax Long Beach CA USA: ACM, Aug. 2023, pp. 4195--4206. [Online]. Available: \url{https://dl.acm.org/doi/10.1145/3580305.3599853}
\BIBentrySTDinterwordspacing

\bibitem{hardesty_5g_2020}
\BIBentryALTinterwordspacing
L.~Hardesty, ``\BIBforeignlanguage{en}{{5G} base stations use a lot more energy than {4G} base stations: {MTN} {\textbar} {Fierce} {Wireless}},'' Apr. 2020, section: Fierce Wireless Homepage,5G,Wireless. [Online]. Available: \url{https://www.fiercewireless.com/tech/5g-base-stations-use-a-lot-more-energy-than-4g-base-stations-says-mtn}
\BIBentrySTDinterwordspacing

\bibitem{sharma_analysis_2021}
\BIBentryALTinterwordspacing
D.~Sharma, S.~Singhal, A.~Rai, and A.~Singh, ``Analysis of power consumption in standalone {5G} network and enhancement in energy efficiency using a novel routing protocol,'' \emph{Sustainable Energy, Grids and Networks}, vol.~26, p. 100427, Jun. 2021. [Online]. Available: \url{https://www.sciencedirect.com/science/article/pii/S2352467720303581}
\BIBentrySTDinterwordspacing

\bibitem{zhou_energy-saving_2013}
\BIBentryALTinterwordspacing
F.~Zhou, J.~Chen, G.~Ma, and Z.~Liu, ``Energy-saving analysis of telecommunication base station with thermosyphon heat exchanger,'' \emph{Energy and Buildings}, vol.~66, pp. 537--544, Nov. 2013. [Online]. Available: \url{https://www.sciencedirect.com/science/article/pii/S0378778813003587}
\BIBentrySTDinterwordspacing

\bibitem{obiora_forecasting_2020}
\BIBentryALTinterwordspacing
C.~N. Obiora, A.~Ali, and A.~N. Hasan, ``Forecasting {Hourly} {Solar} {Irradiance} {Using} {Long} {Short}-{Term} {Memory} ({LSTM}) {Network},'' in \emph{2020 11th {International} {Renewable} {Energy} {Congress} ({IREC})}, Oct. 2020, pp. 1--6, iSSN: 2378-3451. [Online]. Available: \url{https://ieeexplore.ieee.org/abstract/document/9310449}
\BIBentrySTDinterwordspacing

\bibitem{silver_which_2014}
N.~Silver and R.~Fischer-Baum, ``\BIBforeignlanguage{en-US}{Which {City} {Has} {The} {Most} {Unpredictable} {Weather}?}'' Dec. 2014.

\end{thebibliography}

\end{document}